\documentstyle[12pt,fleqn]{article}

\evensidemargin=\oddsidemargin
\begin{document}
\bigskip
\begin{center}
\begin{Large}
\begin{bf}
Theory of the Normal State of Cuprate Superconducting Materials
\end{bf}
\end{Large}
\bigskip
\bigskip
$\;$\\
{\centerline {Yu-Liang Liu}}
International Centre for Theoretical Physics, P. O. Box 586,
34100 Trieste, Italy.\\
\bigskip
{\bf Abstract}
\end{center}
\begin{center}
\begin{minipage}[c]{13cm}
\begin{sf}

\parindent=1.2cm

We have proposed a model Hamiltonian, which describes a simple physical
picture
that the holes with single occupation constraint introduced by doping
move in the antiferromagnetic background of the copper spins,
to describe the normal state of the cuprate superconducting materials,
and used the renormalization group method to calculate its anomalous
magnetic and transport properties. The anomalous magnetic behavior of the
normal state is controlled by both the copper spin and the spin part of
the doping hole residing on the O sites. The physical resistivity is
determined by both the quasiparticle-spin-fluctuation and the
quasiparticle-gauge-fluctuation scatterings and the Hall coefficient is
determined by the parity-odd gauge interaction deriving from the nature
of the hard-core boson which describes the charge part of the doping holes.

\end{sf}
\end{minipage}
\end{center}
\medskip
PACS numbers: 74.20.Mn, 75.10.Jm, 75.40.Gb.
\newpage

Since the discovery of the cuprate superconducting materials$^{[1]}$
there has still considerable controversy over the choice of the
appropriate microscopic Hamiltonian.
Although there have appeared a lot of models, for example, the one-band
effective Hubbard model$^{[2]}$, t-J model$^{[3]}$, three-band Hubbard
model$^{[4]}$, phenomenological marginal Fermi liquid$^{[5]}$, nearly
antiferromagnetic Fermi liquid$^{[6]}$ and so on, to try to describe the
normal and superconducting states of the cuprate superconducting
materials,
it is generally agreed
now that Anderson's starting point$^{[2]}$, namly, strongly on site
Coulomb interactions among a partially filled band of $C{u}$ 3d level,
is correct.
The controversial point is that one is how to treat doping holes residing
on O $2p$ level.
At zero doping, it is generally agreed that the insulating "parent"
phases of the cuprate materials are charge-transfer insulators$^{[7]}$
and can be described by quantum antiferromagnetic Heisenberg model.
Substantial progress has been achieved in understanding of the Heisenberg
limit, both theoretically and experimentally$^{[8-10]}$. Under a finite
doping range, Zhang and Rice$^{[3]}$ showed that the three-band Hubbard
model can be
reduced into a single band effective Hamiltonian--the t-J model under the
case of spin singlet phase that
hybridization strongly binds a hole on each
square of O atoms to the central $C{u}^{2+}$ ions in a similar way as a
hole in the single band effective Hamiltonian, then two holes feel a
strong repulsion against residing on the same square.
In fact, in this representation, the doping hole residing on the O site
only contributes its charge degree and its spin degree is completely
confined to construct the spin singlet with the hole on the Cu site. It
is not clear to what extent it is valid as one uses this representation
to study the effect of the doping.
The gauge theory of t-J model$^{[11-13]}$ gives a good description of the
transport properties of the normal state of the cuprate superconducting
materials, but it is difficult to give a reasonable explanation to its
anomalous magnetic
behavior which is shown by the nuclear magnetic resonance (NMR) and
other experiments.

Recently, it has been shown by Sokol and Pines$^{[14]}$ using purely
scaling considerations, that the experimental data of Refs.[15-18] on the
NMR spin lattice relaxation rate $T_{1}$ and spin echo decay rate
$T_{2G}$ in the cuprates imply a quantum critical(QC), $z=1$ ($z$ a
dynatical exponential), behavior over a broad doping range, and in low
temperature a quantum disorder(QD), $z=1$, behavior. The crossover to the
$z=2$ regime, occurs only in the fully doped materials. This unusual
magnetic behavior reminds one that even in a broad doping range, the
magnetic behavior of the system is still determined by the critical point
there appearing in the zero doping limit. The unusual physical properties
of the normal state may derive from its strongly antiferromagnetic
correlation behavior. The doping will destroy the long range
antiferromagnetic correlation, but the system still remains the short
range antiferromagnetic correlation behavior even in the superconducting
state, which induces the system shifting from renormalized classical(RC)
regime to QC or QD regimes.

It is clear that the remarkably anomalous behavior of the normal state of
the cuprates is mainly on the following aspects in a wide doping range:
a. It shows an antiferromagnetic correlation behavior.
However, the NMR spin-lattice relaxation rates on the Cu site and the O
site in the Cu-O plane have completely different temperature dependence,
the relaxation rate $(T_{1}T)^{-1}$ on the Cu site obeys a Curie-Weiss
law in the higher temperature region, while the relaxation rate
$(^{17}T_{1}T)^{-1}$ on the O site has a linear temperature behavior in
the higher temperature region.
b. It has a linear
temperature dependence of the resistivity in a narrow optimal doping
range, but for the underdoping range, only in the higher temperature
region one has this relation.
c. Its Hall coefficient is
inversely proportional to temperature and the carrier number is of order
doping density. d. It shows a Drude behavior in far-infrared region and
a non-Drude behavior in mid-infrared region.
e. In the underdoping range, there exists a gap (or a pseudogap) in the
spin excitation spectrum$^{[19]}$.
To explain these properties of the normal state but not only one among
them, there needs more effort both on the theory and experiment.
In Ref.20, in contrast to the
t-J model, we loosed the spin degree of the doping hole on the O site and
considered that it has a Kondo-type interaction with the hole on the Cu
site. In fact, in some strongly interacting region, these two models
should give the same physical behavior of the cuprates. Using this model,
we can betterly explain some anomalous magnetic and transport properties
of the normal state of the cuprates. In this paper, we
follow this line to study the normal state more detail.

The common property of the cuprate materials is that they have one or
more Cu-O planes which determine the anomalous behavior of the system in
the normal and superconducting states. In the low temperature region, the
carriers of the system are the doping holes residing on the O sites. It
means that if we want to study the anomalous properties of the system, we
must consider the dynamics of the doping holes, which maybe is a key
point to understand these anomalous behavior. In fact, the real physics
we must consider is that there exists two kinds of the holes, ones reside
on the Cu sites and others on the O sites called the doping holes.
However, the appearence of the mid-infrared region means that there
exists a strongly coupling between the doping holes and the holes on the
Cu sites, but it is not strongly enough to eliminate the spin degree of
the doping hole and enforce one to construct a spin singlet with the hole
on the Cu site at all time. While the doping hole will drastically
influence the magnetic behavior what the holes on the Cu sites have.

 From the current experimental data, we think that the property of
the normal state of the cuprates is determined by two different regions,
one is the central region of the first Brillouin zone of the copper spins,
another is its  corner region, i.e., near the regions ${\bf
Q}=(\pm\frac{\pi}{a},\pm\frac{\pi}{a})$ which mainly reflects the
antiferromagnetic behavior of the system.
We study the following model Hamiltonian$^{[20]}$
\begin{equation}
H=-t\sum_{<ij>}{\tilde{c}}_{i\sigma}^{+}{\tilde{c}}_{j\sigma}
+V_{0}\sum_{i}{\hat{S}}_{i}\cdot{\hat{s}}_{i}
+J\sum_{<ij>}{\hat{S}}_{i}\cdot{\hat{S}}_{j}
\end{equation}
where ${\hat{s}}_{i}=\frac{1}{2}{\tilde{c}}^{+}_{i\alpha}
{\hat{\sigma}}_{\alpha\beta}
{\tilde{c}}_{i\beta}$, ${\tilde{c}}_{i\alpha}=(1-n_{i-\alpha})c_{i\alpha}$,
$c_{i\alpha}(c^{+}_{i\alpha})$
is the hole operator which derives from the doping (i.e., the doping hole
operator). ${\hat{S}}_{i}$ is the spin operator at the $i$ site which
represents
the copper spin. Here we adopt an effective square lattice for O which
identifies the $C{u}$ lattice, the doping holes
have a single occupation
constraint because of the strong repulsive interaction among the doping
holes (we think that in the heavily doping range this constraint is
invalid, it is valid only in the underdoping and optimal doping region).
For more reality, we should consider the doping hole moving in the O
sites, but we take a mapping in phase space, the hopping term can be
written as
$\sum_{p}\epsilon_{p}\tilde{c}^{+}_{p\sigma}\tilde{c}_{p\sigma}$,
$\epsilon_{p}=-4t^{'}cos(\frac{1}{2}ap_{x})cos(\frac{1}{2}ap_{y})$. In
the long wavelength limit, its behavior is the same as that of the first
term in (1).
Second term describes the
Kondo interaction between the copper and doping hole spins which derives
from
the hybridization between the holes on the $C{u}$ sites and the holes on
the $O$ sites.
Last term describes the antiferromagnetic interaction among the copper spins
which is valid for the doping and undoping cases. This model Hamiltonian
can be seen as an effective Hamiltonian deriving from the three bands
Hubbard model where the hybridization between the hole on the $C{u}$
site and the hole on the $O$ site induces the exchange (Kondo)
interaction, here we think that the holes on the $C{u}$ sites are localized,
the doping holes on the $O$ sites are hopping in an effective lattice which
identifies the $C{u}$ sites.  This model Hamiltonian describe a very simple
physical picture
that the doping holes with single occupation condition move in an copper spin
background. In the Kondo regime, the copper and doping hole spins bind into a
Kondo singlet, which is similar to the Zhang and Rice's singlet,
then it has an effective hopping on the different sites. We think that
the model Hamiltonian describes the same physical properties as the t-J model
in some energy scale, but
it can give the NMR data better than the t-J model, and
here the degrees of the doping holes is appearantly
written out.  It is noted that this model is
different from the Kondo
lattice model because of the single occupation condition of the doping holes
and
different from the t-J model because of the appearing of the freedom
degrees of the doping holes.

In fact, this model Hamiltonian is an effevtive Hamiltonian of the
three-band Hubbard model in the strongly on site Coulomb interaction limit.
Because of the strongly on site Coulomb interaction among the holes on the
Cu sites, the copper spin approaches to localization. The doping holes
have strongly hybridization with the holes on the Cu sites, which will
drastically affect the magnetic behavior of the copper spins. In addition
to the strongly on site Coulomb interaction among the doping holes, this
hybridization enforces the charge and spin degrees of the doping holes to
seperate each other. The Kondo-type coupling between its spin part and
the copper spin will determine the magnetic behavior of the system, its
charge part will determine the transport behavior of the system. This
characteristic property of the cuprate superconducting materials derives
from the strongly on site Coulomb interaction and the strong hybridization
between the doping holes and the holes on the Cu sites. But it will
disappear as one heavily dopes the parent insulators. Because of the
doping density increased, the influence of the strongly on site Coulomb
interaction becomes weak and the short-range antiferromagnetic phase of
the system disappears. In some cases, the heavily doping cuprate
superconducting materials should show some characteristic property of the
heavy fermion system if the dimensionality is not important. On the other
hand, if we do not consider the single occupation constraint of the
doping holes, i.e., the doping holes construct a doping conduct band, in
the antiferromagnetic phase of the copper spin, we can easily explain the
magnetic behavior of the normal state$^{[21]}$, but we cannot give a
reasonable explanation to the transport behavior of the normal state.

We adopt the common method to deal with the single occupation condition
by introducing slave boson: ${\tilde{c}}_{i\sigma}=p^{+}_{i}f_{i\sigma}=
b_{i}f_{i\sigma}$,
$p^{+}_{i}p_{i}+f^{+}_{i\sigma}f_{i\sigma}=1$(or $b^{+}_{i}b_{i}=
f^{+}_{i\sigma}f_{i\sigma})$. $b_{i}(=p^{+}_{i})$ is a hard-core boson
operator which describes the charge degree of the hole, $f_{i\sigma}$
is a fermion operator which describes its spin degree. In the representation
of the hole, the Hamiltonian in (1) can be written as
\begin{equation}\begin{array}{rl}
H=& -t\displaystyle{\sum_{<ij>}}(\eta^{*}_{ij}f^{+}_{i\sigma}
f_{j\sigma}+\chi_{ij}b^{+}_{i}b_{j})+V_{0}\displaystyle{\sum_{i}}
b_{i}^{+}b_{i}{\hat{S}}_{i}\cdot{\hat{s}}_{i}\\
+& J\displaystyle{\sum_{<ij>}}{\hat{S}}_{i}\cdot{\hat{S}}_{j}+
t\displaystyle{\sum_{<ij>}}\eta^{*}_{ij}\chi_{ij}+
\displaystyle{\sum_{i}}\lambda_{i}(b^{+}_{i}b_{i}-f^{+}_{i\sigma}
f_{i\sigma})
\end{array}\end{equation}
Here we introduce two Hubbard-Stratonovich fields $\eta^{*}_{ij}$ and
$\chi_{ij}$ to decouple the hard-core boson and fermion, $\lambda_{i}$
is a Lagrangian multiplier which ensures the single occupation condition.
To treat the hard-core nature of the bosons effectively, we make
the following transformation which transforms the hard-core bosons into
the fermions with a vortex tube carrying one flux quantum attached
to each$^{[22,23]}$
\begin{equation}
b_{i}^{+}=h^{+}_{i}exp[-i\sum_{j \not= i}\theta_{ij}n_{j}],
\;\;\;\;\; n_{j}=h^{+}_{j}h_{j}
\end{equation}
where the operators $h^{+}_{i}, h_{i}$ obey Fermi statistics and
$\theta_{ij}$ is the angle between the direction from site $i$ to
site $j$ and some fixed direction, the x axis for example. We think
that in the low temperature, low energy and long wavelength limits,
the hard-core nature of the bosons is important, we must consider
its effect.

In the spin and fermion coherent state representations,
for spin part, we take the antiferromagnetic N\'{e}el order as its
background, because although the doping destroies the long range
antiferromagnetic order, the short range antiferromagnetic order is still
remained.
The reason of taking this approximation is that at zero doping, the
magnetic behavior of the system is controlled by a quantum critical point
which is determined by the coupling constant and temperature
parameters$^{[8]}$. The doping will drastically influence the magnetic
behavior of the system, but it doe not change the characteristic property
of the quantum critical point which still determines the magnetic behavior
of the doping system. Because of the interlayer very weak interaction,
there exists a long range antiferromagnetic N\'{e}el order at low
temperature for the parent insulators. Even in the fully (optimal) doping
range, the doping system still has the short-range antiferromagnetic
order both in the normal state and in the superconducting state, so the
antiferromagnetic
behavior is one of the remarkable character properties of the cuprate
materials. Therefore it is reasonable that we take this approximation as
a starting point to study the magnetic behavior of the cuprate materials.
But in the heavily doping range, this starting point will be wrong if the
short-range antiferromagnetic order disappears. For the doping hole
part, we
take the following approximations, i.e., only consider the effect of the
phase fluctuation
\begin{equation}\begin{array}{rl}
\eta_{ij}=& <\eta>e^{iA_{ij}},\;\;\;    \chi_{ij}= <\chi>e^{iA_{ij}}\\
\lambda_{i}=& \lambda+iA_{0}(i), \;\;\;   A_{ij}= ({\bf r}_{i}-{\bf
r}_{j})\cdot {\bf A}(\frac{{\bf r}_{i}-{\bf r}_{j}}{2})
\end{array}\end{equation}
It is well known that the Hamiltonian (2) can be transformed into as
following action
\begin{equation}\begin{array}{rl}
S[a_{\mu},\psi_{h}]=& \displaystyle
{\frac{1}{2g}\int^{\beta}_{0}}d\tau\int d^{2}x[({\bf \nabla}\hat{\Omega})^{2}
+\frac{1}{c^{2}}(\partial_{\tau}\hat{\Omega})^{2}]\\
+& \displaystyle
{ \int^{\beta}_{0}}d\tau\int d^{2}x\{\psi^{*}_{f\sigma}(\partial_{\tau}-
\mu_{F}-ieA_{0})\psi_{f\sigma}\\
+& \psi^{*}_{h}(\partial_{\tau}-\mu_{B}+ieA_{0}
+iea_{0})\psi_{h}\\
-& \displaystyle
{\frac{1}{2m_{F}}}\psi^{*}_{f\sigma}({\bf \nabla}-ie{\bf
A})^{2}\psi_{f\sigma}
-\frac{1}{2m_{B}}\psi^{*}_{h}({\bf \nabla}+ie{\bf a}+ie{\bf
A})^{2}\psi_{h}\}\\
+&  \displaystyle{\sum_{n}\int\frac{d^{2}q}{(2\pi)^{2}}}
u\hat{s}(q,\omega_{n})\cdot\hat{\Omega}(Q-q,-\omega_{n})\\
+& \displaystyle{ \frac{\alpha}{4i\pi}\int^{\beta}_{0}d\tau\int d^{2}x}
\varepsilon_{\mu\nu\lambda}a_{\mu}\partial_{\nu}a_{\lambda}
\end{array}\end{equation}
where $\alpha=\frac{e^{2}}{2l+1}, l=0,1,2,...$; ${\bf
Q}=(\pm\frac{\pi}{a},\pm\frac{\pi}{a}), u=V_{0}S\delta,
g=\frac{1}{JS^{2}}, c^{2}=8a^{2}J^{2}S^{2}$, $m_{F}=\frac{1}{t<\eta>}$,
$m_{B}=\frac{1}{t<\chi>}$, $\delta$ is the doping density,
$\hat{\Omega}(x_{i})=\eta_{i}\hat{S}_{i}/S, \eta_{i}=\pm$, is the staggered
spin field, $\psi_{f\sigma}, \psi_{h}$ the fermion fields,
$A_{\mu}$ the gauge field which derives
from the phase fluctuation of the Hubbard-Stratonovich fields $\eta_{ij}$
and $\chi_{ij}$,
$a_{\mu}$ the Chern-Simons gauge field which derives from the nature of the
hard-core boson.
If we integrate out the gauge field $a_{\mu}$, it is equivalent to
take the following transformation in the action
\begin{equation}
\int Da_{\mu}D\psi_{h}e^{-S[a_{\mu},\psi_{h}]}=\int D\phi e^{-
S[0,\phi]}
\end{equation}
where $\phi$ is the hard-core boson field.

We can easily see that the physical picture described by the action in (5)
is more clear, the spin and charge degrees of the doping hole are separated,
but there exists an interaction between them via the gauge field.
The spin part of the doping hole has an interaction with the staggered spin
field,
which mainly describes the magnetic behavior of the normal state. The
transport properties of the normal state is dominanted by both the charge
part of the doping hole and the quasiparticle-spin-fluctuation scattering.
The spin part of the system is
mainly controlled by the coner region of the first Brillouin zone ${\bf
q}\sim {\bf Q}$, but
the charge part of the system is mainly controlled by its central region
${\bf q}\sim 0$.
We think that the spin part of the system does not drastically affect its
charge part and will not change its transport properties at the normal
state if the quasiparticle-spin-fluctuation scattering is not
dominant, because
the staggered spin field has less influence on the hard-core
boson field and gauge field. But in the superconducting state, the spin
part does drastically affect the charge part. The antiferromagnetic spin
fluctuation tends to make the fermions pair and destroy the gauge invariance,
but the gauge field is strongly pair breaking, and will in general supress
the pairing transition temperature significantly. The competition of the
antiferromagnetic fluctuation and gauge field fluctuation will
determine the transition temperature of the superconducting state.
This problem will be addressed in a seperate paper. Here we only
consider the physical properties of the normal state.

Generally, at the normal state, we can integrate out the fermion field
$\psi_{f\sigma}$ and obtain an effective action including the staggered spin
field, hard-core boson field and gauge field. Because of the gauge invariance,
the staggered spin field does not directly interact with the gauge field,
they only affect each other by through the fermion field. After integrating
out the fermion field, we have the following effective action
\begin{equation}
S_{eff.}=S^{s}_{eff.}+S^{c}_{eff.}
\end{equation}
\begin{equation}
S^{s}_{eff.}=\beta\sum_{n}\int\frac{d^{2}q}{(2\pi)^{2}}\{\frac{1}{2g}
(q^{2}+\frac{1}{c^{2}}\omega^{2}_{n})|\hat{\Omega}|^{2}-
F(Q-q)\frac{|\omega_{n}|}{\omega_{AF}}|\hat{\Omega}|^{2}\}
\end{equation}

\begin{equation}\begin{array}{rl}
S^{c}_{eff.}=& \displaystyle
{ \int^{\beta}_{0}}d\tau\int d^{2}x\{ \psi^{*}_{h}(\partial_{\tau}
-\mu_{B}+iea_{0}+ieA_{0})\psi_{h}\\
-& \displaystyle
{\frac{1}{2m_{B}}}\psi^{*}_{h}({\bf \nabla}+ie{\bf a}+ie{\bf A})^{2}
\psi_{h}\}\\
+& \displaystyle
{ \sum_{\omega_{n},q}}[\chi_{F}q^{2}+\frac{
\gamma_{1}|\omega_{n}|}{q}]
\cdot\displaystyle
{ [\delta_{ij}-\frac{q_{i}q_{j}}{q^{2}}]}
A_{i}(q,i\omega_{n})A_{j}(-q,-i\omega_{n})\\
+& \displaystyle{\frac{\alpha}{4i\pi}\int^{\beta}_{0}\int
d^{2}x} \varepsilon_{\mu\nu\lambda}a_{\mu}\partial_{\nu}a_{\lambda}
\end{array}\end{equation}
where, $\omega_{AF}=\frac{2v_{F}}{aN(E_{F})}\frac{1}{u^{2}}$,
$\gamma_{1}=\frac{\pi}{v_{F}}$,
$N(E_{F})$ is the density of the state at the Fermi surface, $\chi_{F}$
and $v_{F}$ are the diamagnetic susceptibility and the Fermi wave velocity,
the factor $F(Q-q)$ can be written in the
one-loop approximation $F(Q-q)\sim\frac{\pi}{a|{\bf Q}-{\bf q}|}\sim 1$ for
$q<<1$.
We have omited term
$A_{0}(q,i\omega)\bar{D}_{00}(q,i\omega)A_{0}(-q,-i\omega)$, because of
longitudinal screening effect, in the limit of the low energy and long
wavelength, $\bar{D}_{00}(q,i\omega)$ takes constant value,
$\bar{D}_{00}(q,i\omega)\longrightarrow const. \;\; as\;\;
\omega,q\longrightarrow 0$. We see
that the spin and charge parts are completely separated (here omiting
the high order term). The higher terms may induce the interaction between
the spin and charge parts, but they are the irrelevant terms in the low
energy limit.
This is true because the Fermi surface is stable
which is not destroyed by the gauge fluctuation due to the
nature of the hard-core bosons,
one can safely integrate out the fermion field.

Now we separately study the spin and charge parts.
We can use the renormalization group method to study the spin
part$^{[8,24]}$.
 From the action (8), we see that there exist three regimes,
$z=1$ ($z$ a dynamic exponential) regime,
the square term of frequency $\omega$ is dominant over its linear term;
$z=2$ regime,
the linear term is dominant;
and the crossover regime
from $z=1$ to $z=2$.
The $z=1$ and $z=2$ regimes are controlled by different quantum critical
points, the $z=1$ regime is controlled by the quantum critical point of
the undoping parent insulators, the $z=2$ regime is controlled by a new
quantum critical point induced by the overdamping spin wave effect which
derives from the coupling between the copper spins and the doping holes.
The crossover regime is a border of these two regimes. This
classification depends upon the assumption that the short-range
antiferromagnetic order is still robust under the doping. If the
short-range antiferromagnetic order of the system disappears, one cannot
use the action (8) to describe the system. There exists a character
energy $\omega_{s}$, as $\omega_{AF}\gg\omega_{s}$, the overdamping
effect of the spin spectrum is less important, the magnetic behavior of
the system is determined by the $z=1$ regime, as
$\omega_{AF}\ll\omega_{s}$, the overdamping effect is dominant, the
magnetic behavior of the system is determined by the $z=2$ regime.

We use the renormalization group methods developed in Refs.[8,24] to
study the behavior of the effective action in (8). Here we
adopt the symbols in Ref.8, $g_{0}=hc\Lambda g, t_{0}=k_{B}Tg, \Lambda$
is a cutoff of the wave vector. For intermediate doping, the last term
in (8) is a small quantity which can be treated perturbatively,
the frequency $\omega$ has the scaling transformation
$\omega^{\prime}=\omega e^{l}$ which corresponds to the z=1 regime.
In order to get the low energy behavior of the system, we can integrate
out the high energy parts which will induce the effective coupling
constants depending upon the renormalization parameter $l$.
In one-loop approximation, we can get the following renormalization
group equations of the coupling constants
\begin{equation}\begin{array}{rl}
\displaystyle{\frac{dt}{dl}}=& \displaystyle{\frac{gt}{4\pi}
\frac{1}{\sqrt{1-a^{2}g^{2}}}}\displaystyle{\frac{sinh(\frac{g}{t}
\sqrt{1-a^{2}g^{2}})}{cosh(\frac{g}{t}\sqrt{1-a^{2}g^{2}})-
cos(\frac{ag^{2}}{t})}}\\

\displaystyle{\frac{da}{dl}}=& [2-
\displaystyle{\frac{g}{4\pi}
\frac{1}{\sqrt{1-a^{2}g^{2}}}}\displaystyle{\frac{sinh(\frac{g}{t}
\sqrt{1-a^{2}g^{2}})}{cosh(\frac{g}{t}\sqrt{1-a^{2}g^{2}})-
cos(\frac{ag^{2}}{t})}}]a\\

\displaystyle{\frac{d}{dl}(\frac{g}{t})}& = -\displaystyle{\frac{g}{t}}

\end{array}\end{equation}
where $a_{0}=\frac{1}{\omega_{AF}}$. If we assume the density of state
at the Fermi surface $N(E_{F})\propto m_{F}$, the quantity $\omega_{AF}$
takes the form $\omega_{AF}\propto\frac{1}{N^{2}(E_{F})}\frac{1}{u^{2}}$.
For simplicity, we can omit the dependence
of the first equation on $a$, and only keep the linear term in the second
equation in (10). These approximations will be enough to give the correct
behavior of the
characteristic frequency $\omega_{AF}$ and correlation length $\xi$ in
different regimes (see below). With the above approximations,
we have
\begin{equation}\begin{array}{rl}\displaystyle
t(l)=&
\displaystyle{\frac{t_{0}}
{1+\displaystyle{\frac{t_{0}}{2\pi}ln\frac{sinh(\frac{1}{2}x_{0}e^{-l})}
{sinh(\frac{1}{2}x_{0})}}}}\\
g(l)=& \displaystyle{\frac{g_{0}e^{-l}}
{1+\displaystyle{\frac{t_{0}}{2\pi}ln\frac{sinh(\frac{1}{2}x_{0}e^{-l})}
{sinh(\frac{1}{2}x_{0})}}}}\\
a(l)=& \displaystyle{a_{0}e^{2l}\frac{t_{0}}{t(l)}}
\end{array}\end{equation}
where $x_{0}=\frac{g_{0}}{t_{0}}$. In the QC regime, we have
($t(\hat{l})=2\pi$)$^{[8]}$
\begin{equation}\begin{array}{rl}
\xi=& a\displaystyle{e^{\hat{l}}\sim\frac{1}{T}}\\

a(\hat{l})=&
\displaystyle{a_{0}(\frac{\xi}{a})^{2}\frac{k_{B}T}{2\pi\rho^{0}_{s}}}
\end{array}\end{equation}
In the QD regime, we have ($g(\hat{l})=8\pi$)
\begin{equation}\begin{array}{rl}
\displaystyle
\xi=& const.\\
a(\hat{l})=& \displaystyle{\frac{1}{8\pi}a_{0}g_{0}(\frac{\xi}{a})}
\end{array}\end{equation}
Here we see that the correlation length $\xi$ given in (12) and (13) is
independent of the doping
density $\delta$. If one solves equations (10) without using the above
approximations, one can find that the correlation length $\xi$ will
decrease with increasing the doping density $\delta$ because of the
quantity $a_{0}$ depending upon the doping density $\delta$. In the present
case, the
RC regime will disappear, because in the intermediate doping range, the
system enters the QC or QD regime.

On the other hand, for fully doping the last term in (8) is
more important than the second term, we can not treat it as a perturbative
term, the frequency $\omega$ has the scaling transformation
$\omega^{\prime}=\omega e^{2l}$, corresponding to the $z=2$ regime,
which means that the system undergoes the crossover regime from
the $z=1$ to the $z=2$ regimes due to the doping. In the $z=2$ regime,
the second term in (8) and high order terms of the frequency are
irrelevant under the scaling transformation, we will omit these terms.
Using the above methods, in one-loop approximation we have
\begin{equation}\begin{array}{rl}
\displaystyle{\frac{dt}{dl}}=& \displaystyle{\frac{gt}{4\pi}
ctg(\frac{g}{2t}\overline{\omega})}\\
\displaystyle{\frac{d}{dl}(\frac{g}{t})}& =  -2\displaystyle{\frac{g}{t}}
\end{array}\end{equation}
here $\bar{\omega}=\frac{\omega_{AF}}{2g_{0}}$ will be treated as a
renormalization invariance.
We can easily solve the equations (14) and get following expressions
\begin{equation}\begin{array}{rl}
t(l)=& \displaystyle{\frac{t_{0}}{1+
\displaystyle{\frac{t_{0}}{4\pi\bar{\omega}} ln\frac{sin(\frac{1}{2}x_{0}
\bar{\omega}e^{-2l})}{sin(\frac{1}{2}x_{0}\bar{\omega})}}}}\\
g(l)=&  \displaystyle{\frac{g_{0}e^{-2l}}{1+
\displaystyle{\frac{t_{0}}{4\pi\bar{\omega}}
ln\frac{sin(\frac{1}{2}x_{0}\bar{\omega}e^{-2l})}
{sin(\frac{1}{2}x_{0}\bar{\omega})}}}}
\end{array}\end{equation}
In the QC regime, we have ($t(l)=2\pi$)
\begin{equation}
\xi^{'2}\sim\frac{1}{\bar{\omega}T}
\end{equation}
In the QD regime, we have ($g(l)=8\pi$)
\begin{equation}
\xi^{'}\sim const.
\end{equation}
We see that there exists qualitatively difference between the $z=1$ and
$z=2$ regimes. In the $z=1$ regime, the correlation length $\xi$ is
inversely proportional to the temperature $T$, and the characteristic
energy $\omega_{AF}$ is renormalized. However, in the $z=2$ regime, the
square of the correlation length is inversely proportional to the
temperature, $\xi^{'2}\sim\frac{1}{T}$, because the characteristic energy
$\omega_{AF}$ is very less than the character energy $\omega_{s}$, it is
not renormalized.
In fact, equation (10) and (14) is obtained in the different critical
regions, they are valid only near their fixed points correspondence with
the $z=1$ and $z=2$ regions, respectively. We can intuitively understand
these problem in this way, in the low doping, the density of state
$N(E_{F})$ is very small, then the quantity $\omega_{AF}$ is very larger
than the character energy $\omega_{s}$,
the linear term of $\omega_{n}$ is less important than the quadratic term
of $\omega_{n}$, the system is in the $z=1$ region. In the large doping,
the quatity $\omega_{AF}$ is very less than the character energy
$\omega_{s}$, the linear term of $\omega_{n}$
is more important than the quadratic term, the system is going into the
$z=2$ region.

Generally, for the $z=1$ regime, according to these resolutions
we can write the following
expression of the spin susceptibility in the low energy limit
\begin{equation}
\chi_{1}(q,\omega)=\frac{\chi_{0}}{\xi^{-2}+q^{2}-\frac{1}{c^{2}}\omega^{2}
-iF(Q-q)\frac{\omega}{\omega^{R}_{AF}}}
\end{equation}
where, $\frac{1}{\omega^{R}_{AF}}=\frac{4\pi a^{2}}{k_{B}T\omega_{AF}(\hat{l})
\xi^{2}}+\frac{1}{\omega_{R}}$, $\xi\sim\frac{1}{T}$, for QC regime;
$\frac{1}{\omega^{R}_{AF}}=\frac{16\pi a}{\omega_{AF}(\hat{l})\xi}+
\frac{1}{\bar{\omega}_{R}(T)}$,
$\xi\sim const.$,
for QD regime. $\omega_{R}$ and $\bar{\omega}_{R}$
derive from the high order quantum fluctuation without the doping$^{[10]}$,
in QD regime, $\bar{\omega}_{R}$ is very large, as $T\rightarrow
0, \bar{\omega}_{R}(T)\rightarrow\infty$; In QC regime, $\omega_{R}\sim\lambda
T(\frac{\xi}{a})^{2}$, $\lambda$ is a constant. $\omega_{AF}(\hat{l})$
is the renormalized quantity of $\omega_{AF}$.
We see that the image term is consist of two parts, one is deriving
from the effect of undoping and another is induced by doping.
For the $z=2$ regime, we can
write a general expression of the spin susceptibility in the low energy limit
\begin{equation}
\chi_{2}(q,\omega)=\frac{\chi'_{0}}{\xi^{'-2}+q^{2}-iF(Q-q)\frac{\omega}
{\bar{\omega}_{AF}}}
\end{equation}
where,
$\xi^{'2}\sim\frac{1}{T}$, in QC regime; $\xi^{'}\sim const.$, in QD
regime.
We need some explanation for the quantities $\chi_{0}$ and
$\chi^{'}_{0}$. In the QD regimes of the $z=1$ and $z=2$ regimes, because
of the correlation length taking constant values, there will appear a gap
$\Delta_{0} (\Delta^{'}_{0})$ in the spin spectrum, the spin wave
excitation will be suppressed, so the quantities $\chi_{0}$ and
$\chi^{'}_{0}$ should be exponentially decaying functions as decreasing
the temperature in the QD regimes. In the QC regimes, the spin wave
excitation energy is very larger than the energy gap(s) $\Delta_{0}
(\Delta^{'}_{0})$, so one can take the quantities $\chi_{0}$ and
$\chi^{'}_{0}$ as constants.
These two expressions of the spin susceptibility
are valid in the coner region
and mainly describe the physical
properties of the normal state controlled by the coner region (i.e.,
${\bf Q}=(\pm\frac{\pi}{a},\pm\frac{\pi}{a}))$ of the first Brillouin zone.

Using above spin susceptibilities, we can calculate the NMR spin lattice
relaxation rate $T_{1}$ and spin echo rate $T_{2G}$
at $C{u}$ sites which are completly
determined by the real and imagenary parts of the spin susceptibility,
respectively.
In the QD regime we have
\begin{equation}\begin{array}{rl}
\displaystyle
\frac{1}{T_{1}T}& \propto\left\{
\begin{array}{ll}\chi_{0},
&\mbox{z=1}\\
\chi^{'}_{0},
&\mbox{z=2}\end{array}\right.\\
\displaystyle{\frac{1}{T_{2G}}}& \propto\left\{
\begin{array}{ll}\chi_{0},
&\mbox{z=1}\\
\chi^{'}_{0},
&\mbox{z=2}\end{array}\right.
\end{array}\end{equation}
if the hyperfine coupling constants for both the relaxation rate and the
spin
echo rate have the similar momentum dependence. Similarly, in the QC
regime we have
\begin{equation}\begin{array}{rl}
\displaystyle
\frac{1}{T_{1}T}& \propto\left\{
\begin{array}{ll}\frac{1}{\omega_{AF}(\hat{l})T}
+\frac{1}{\omega_{R} T},
&\mbox{z=1}\\
\frac{1}{{\bar{\omega}}^{2}_{AF}T},
&\mbox{z=2}\end{array}\right.\\
\displaystyle{
\frac{1}{T_{2}}}& \propto\left\{
\begin{array}{ll}\frac{1}{T},
&\mbox{z=1}\\
\frac{1}{{\bar{\omega}_{AF}T}^{1/2}},
&\mbox{z=2}\end{array}\right.
\end{array}\end{equation}
We see that the NMR relaxation rate $\frac{1}{TT_{1}}$ increases and
reaches a top point and then decreases as the temperature $T$ increases.

Now we consider the NMR spin-lattice relaxation rate on O site which is
mainly determined by the ${\bf q}\sim 0$ region,
$(^{17}T_{1}T)^{-1}\propto\chi(q=0)$, $\chi(q=0)$ being a static spin
susceptibility. If we take the relation for the static spin
susceptibility$^{[25]}$ $\chi(q=0)\propto N_{R}(E_{F})$, $N_{R}(E_{F})$
being a renormalized density of state, we can explain the temperature
dependence of the static spin susceptibility and the relaxation rate on O
site. In equation (5), the spin coupling term is
\[
\frac{1}{2} u
\sum_{n}\int\frac{d^{2}q}{(2\pi)^{2}}\frac{d^{2}q^{'}}
{(2\pi)^{2}}d\omega\psi^{*}_{f\alpha}(q,\omega)\hat{\sigma}_{\alpha\beta}
\psi_{f\beta}(q+q',\omega+\omega_{n})\cdot\hat{\Omega}(Q-q',-\omega_{n})
\]
If we adopt the method in Ref.26 to make the renormalization group
transformation for the fermion field $\psi_{f\sigma}$, we can obtain the
relation $u(l)\propto ue^{2l}$, according to the equation (10),
$a(l)\propto (N(E_{F})u)^{2}(l)\propto a_{0}e^{2l}$, so we have the
relations $N(E_{F},l)\propto N(E_{F})e^{-l}$,
$N_{R}(E_{F})=N(E_{F},\hat{l})\propto N(E_{F})/\xi$. If the system is in
the $z=1$ QC regime, $\xi\propto\frac{1}{T}$, we have the relation
\begin{equation}
(^{17}T_{1}T)^{-1}\propto\chi(q=0)\propto T
\end{equation}
which is reasonable at least in the ${\bf q}\sim 0$ regime. If the system
is in the $z=2$ QC regime, using the same method as above, we can have
the relation $\chi(q=0)\propto\sqrt{T}$. However, in the $z=2$ QC regime,
the doping holes can construct a big Fermi surface, quasiparticle-hole
pairing excitation heavily dampes the spin wave spectrum. If we assume
that the main process of the quasiparticle-hole pairing excitation is the
quasiparticle-hole pair $({\bf q}, {\bf Q+q})$, the spin coupling term in
(5) can be written as
\[
\frac{1}{2}u\sum_{n}\int\frac{d^{2}q}{(2\pi)^{2}}\frac{d^{2}q'}{(2\pi)^{2}}
d\omega\psi^{*}_{f\alpha}(q,\omega)\hat{\sigma}_{\alpha\beta}\psi_{f\beta}
(Q+q+q', \omega+\omega_{n})\cdot\hat{\Omega}(-q', -\omega_{n})
\]
Using the same method as above, we find the coupling constant
$u(l)=ue^{l}$, so we have the relation
\begin{equation}
(^{17}T_{1}T)^{-1}\propto\chi(q=0)\sim const.
\end{equation}
which is correct at zeroth order approximation, the higher orders
contribute a small quantity. However, in Ref.27, the authors showed that
the higher order term can give a correction $-\alpha T$, $\alpha\ll 1$,
to the spin susceptibility $\chi(q=0)$ in (23) for a large Fermi surface.

In the low temperature region,
the system entres into the QD regime, there will appear the
energy gap $\Delta_{0}$ in the antiferromagnetic quantum fluctuation
excitation spectrum in the ${\bf q}\sim{\bf Q}$ region.
However, the opening energy gap also drastically
influences
the static spin susceptibility $\chi(q=0)$ and the NMR relaxation
rate at $O$ sites.
We think that the relation $\frac{1}{^{17}T_{1}T}\propto\chi(q=0)$ can be
approximately extended to the QD regime and the static spin
susceptibility $\chi(q=0)$ is an exponentially decaying function as
decreasing the temperature if the energy gap in the
spin excitation spectrum extendes over the whole Fermi surface.
These results are in good agreement with the current experimental
data which shows that the model Hamiltonian in (1) can be betterly
used to describe the magnetic behavior of the normal state of the
cuprates. We think that the model Hamiltonian captures the key point
of the cuprate superconducting materials that the holes induced by the
doping has a strongly magnetic correlation with the copper spin,
this property is responsible for the anomalously magnetic behavior
of the cuprate superconducting materials.

At the normal state, the transport property of the system is mainly
determined by the effective action (9) and the fermion $\psi_{f\sigma}$.
First we consider the contribution from the fermion $\psi_{f\sigma}$.
Using the expression given in Ref.6, we study the behavior of the
resistivity produced by the quasiparticle-spin-fluctuation scattering
\begin{equation}\begin{array}{rl}
\rho_{\psi}(T)\propto & \displaystyle{\frac{1}{T}\int d\omega d^{2}q
\frac{\omega e^{\omega/T}}{(e^{\omega/T}-1)^{2}}\chi^{"}(q,\omega)}\\
\propto & T\displaystyle{\int^{\infty}_{0}dx\frac{xe^{x}}{(e^{x}-1)^{2}}
tan^{-1}(\frac{T\xi^{2}_{i}x}{\omega_{i}[1-\alpha_{i}(\xi^{2}_{i}
T^{2}x^{2}/c^{2})]})}
\end{array}\end{equation}
where $\alpha_{i}=0\; for\;z=2\;or\;1\;for\;z=1$.
$\xi_{i}=\xi^{'}\;for\;z=2\;or\;\xi\;for\;z=1$,
$\omega_{i}=\bar{\omega}_{AF}\;for\;z=2\;or\;\omega^{R}_{AF}\;for\;z=1$.
We see that, in the QC ($z=1\;or\;2$) regime, the resistivity varies
linearly with the temperature $\rho_{\psi}(T)\propto T$, and its slope
depends upon the doping density. However, in the QD regime, the
resistivity can be nearly written as $\rho_{\psi}(T)\propto T^{\alpha}$,
i.e.,
\begin{equation}
\rho_{\psi}(T)\propto \left \{ \begin{array}{ll}
T,  &\mbox{$T>T^{*}$}\\
T^{\alpha},& \mbox{$T<T^{*}$}
\end{array}\right.
\end{equation}
where $\alpha\sim 2,\; for\;T\ll T^{*}$, $T^{*}$ is a characteristic
temperature indicating the system going from the ($z=1\;or\;2$) QC regime
into the QD regime. While as the system goes from the $z=2$ QC regime
into the $z=1$ QC regime, the resistivity still varies linearly with the
temperature, but its slope will be changed. The resistivity produced by
the quasiparticle-gauge-fluctuation scattering is$^{[12]}$
$\rho_{F}^{'}(T)\propto T^{4/3}$. The physical resistivity should be
\begin{equation}
\rho(T)=\rho_{\psi}(T)+\bar{\rho}(T)
\end{equation}
where $\bar{\rho}(T)=\rho_{F}(T)+\rho_{B}(T)\sim\rho_{B}(T)$,
$\rho_{B}(T)$ is the contribution of the hard-core boson.

However, so far there is not an
effective method to deal with the hard-core boson, although one often
meets it in the literature. Here we follow the method in Ref.20 to deal with
the hard-core boson. In a higher energy range, the hard-core boson shows the
behavior of a
boson; In a lower energy range, it effectively shows the nature of a
fermion. So there exists a character energy $\omega_{c}$, as
$\omega\gg\omega_{c}$, we can treat it as a boson, as
$\omega\ll\omega_{c}$, we should treat it as a fermion. For the former
case, it has been extensively studied by many authors$^{[11-13]}$. In
this case, the action (9) can be written as
\begin{equation}\begin{array}{rl}
S^{c}_{eff.}=& \displaystyle
{ \int^{\beta}_{0}}d\tau\int d^{2}x\phi^{*}[\partial_{\tau}
+ieA_{0}
+\displaystyle
{\frac{1}{2m}}({\bf \nabla}+ie{\bf A})^{2}]\phi\\
+&  \displaystyle{\sum_{\omega_{n},q}(\chi_{F}q^{2}+\frac{\gamma_{1}
|\omega_{n}|}{q})(\delta_{ij}-\frac{q_{i}q_{j}}{q^{2}})}
A_{i}(q,i\omega_{n})A_{j}(-q,-i\omega_{n})
\end{array}\end{equation}
where $\phi$ is the boson field.
It shows the resistivity $\rho_{B}(T)$ has
a linear temperature dependence and the Hall coefficient is inversely
proportional to the doping density (holon density), but it cannot explain
the temperature dependence of the Hall coefficient which we think that it
derives from the nature of the hard-core boson.

To determine the low energy and long wavelength behavior of the gauge
field, we integrate out the fermion field $\psi_{h}$ in the action (9) and
obtain an effective action
\begin{equation}\begin{array}{rl}
S_{eff.}=&
\displaystyle{\sum_{\omega_{n},q}(\chi_{F}q^{2}+\frac{\gamma_{1}
|\omega_{n}|}{q})(\delta_{ij}-\frac{q_{i}q_{j}}{q^{2}})}
A_{i}(q,i\omega_{n})A_{j}(-q,-i\omega_{n})\\
+&  \displaystyle{\sum_{\omega_{n},q}(\chi q^{2}+\frac{\gamma_{2}
|\omega_{n}|}{q})(\delta_{ij}-\frac{q_{i}q_{j}}{q^{2}})}
a_{i}(q,i\omega_{n})a_{j}(-q,-i\omega_{n})\\
+&  \displaystyle{\sum_{\omega_{n},q}}
a_{0}(q,i\omega_{n})D_{00}(q,i\omega_{n})a_{0}(-q,-i\omega_{n})\\
-& \displaystyle{\frac{\alpha}{4\pi}\sum_{\omega_{n},q}}\epsilon_{\mu\nu
\lambda}(a-A)_{\mu}q_{\nu}(a-A)_{\lambda}
\end{array}\end{equation}
where $q_{\mu}=(q_{i},\omega_{n})$, we have taken the Coulomb gauge
${\bf \nabla}\cdot{\bf a}={\bf \nabla}\cdot{\bf A}=0$.
If there is not longitudinal screening effect for the CS gauge field,
$D_{00}(q,i\omega)$ will take the value $D_{00}(q,i\omega)\sim q^{2}$; If
there is longitudinal screening effect, $D_{00}(q,i\omega)$ would take
the constant value in the limit of low energy and long wavelength,
$D_{00}(q,i\omega)\sim const.$. The longitudial screening effect of the
CS gauge fields will drastically influence the low energy behavior of the
system. First, we consider the latter case, there existing the
longitudinal screening effect of the CS gauge field. Because of
$D_{00}(q,i\omega)$ taking constant value, the CS term in (28) cannot
provide a gap to the transverse parts of the gauge field and CS gauge
field. So we can use the method in Ref.24 and take following scaling
transformations
\begin{equation}
q\rightarrow sq, \;\; \omega_{n}\rightarrow s^{3}\omega_{n},
\;\; s\rightarrow 0
\end{equation}
We see that the dynamic exponent is $z=3$. We must notice that these (and
below) scaling transformations are taken in the ${\bf q}\sim 0$
(not ${\bf q}\sim {\bf Q}$) region of the first Brillouin zone. To keep the
quadratic
term invariance, the gauge field and CS gauge field would take following
scaling transformations
\begin{equation}\begin{array}{rl}
a_{0}& \longrightarrow \displaystyle{s^{-5/2}}a_{0}, \;\;
a_{i}\longrightarrow\displaystyle{s^{-7/2}}a_{i}\\
A_{0}& \longrightarrow\displaystyle{s^{-5/2}}A_{0}, \;\;
A_{i}\longrightarrow\displaystyle{s^{-7/2}}A_{i}
\end{array}\end{equation}
Under these transformations, all higher order interaction vertex functions
$\Gamma^{(n)}A^{n}$ and $\Gamma^{'(n)}a^{n}, n\geq 3$, are
irrelevant$^{[28-30]}$. Let us see
the CS term, the terms $\sum\epsilon_{ij}\omega_{n}a_{i}(a-A)_{j}$ and
$\sum\epsilon_{ij}\omega_{n}A_{i}(a-A)_{j}$ are irrelevant because of their
scaling dimension being one, the terms
$\sum\epsilon_{ij}q_{i}a_{0}(a-A)_{j}$
and $\sum\epsilon_{ij}q_{i}A_{0}(a-A)_{j}$ are marginal because of their
scaling dimension being zero.
We see that in $z=3$ regime, the coupling constant $e$ and statistical
parameter $\alpha$ are exactly marginal. The physical property of the
system is controlled by following effective action at the quantum
critical point $e(s)=e(0)$ and $\alpha(s)=\alpha(0)$
\begin{equation}\begin{array}{rl}
S^{c}_{eff.}=&
\displaystyle{\sum_{\omega_{n},q}(\chi_{F}q^{2}+\frac{\gamma_{1}
|\omega_{n}|}{q})(\delta_{ij}-\frac{q_{i}q_{j}}{q^{2}})}
A_{i}(q,i\omega_{n})A_{j}(-q,-i\omega_{n})\\
+&  \displaystyle{\sum_{\omega_{n},q}(\chi q^{2}+\frac{\gamma_{2}
|\omega_{n}|}{q})(\delta_{ij}-\frac{q_{i}q_{j}}{q^{2}})}
a_{i}(q,i\omega_{n})a_{j}(-q,-i\omega_{n})\\
+&  \displaystyle{\sum_{\omega_{n},q}}
a_{0}(q,i\omega_{n})D_{00}(q,i\omega_{n})a_{0}(-q,-i\omega_{n})\\
-&  \displaystyle{\frac{\alpha}{4\pi}\sum_{\omega_{n},q}}\epsilon_{ij}q_{i}
(a-A)_{0}(q,i\omega_{n})(a-A)_{j}(-q,-i\omega_{n})
\end{array}\end{equation}
We see that the parity-odd term $\epsilon_{ij}q_{i}a_{0}a_{j}$
survives. This term will determine the anomalous behavior of the
Hall coefficient (see below).

Using the gauge propagators given in (31),
by a simple calculation, the imaginary parts of the
fermion self-energy and its Green function can be written as,
respectively$^{[28-30]}$
\begin{equation}
\Sigma^{"}_{c}(k_{F},\omega)\sim \omega^{2/3}, \;\;\;\;  G^{"}_{c}(k_{F},
\omega)\sim\omega^{-2/3}
\end{equation}
We see that at the quantum critical point, the spectral density
$G^{"}_{c}$ has a power law divergence, removing all remnant characters of
the quasiparticle and destroying the Fermi liquid behavior.

If we consider that there is not longitudinal screening effect for the CS
gauge field, i.e., $D_{00}(q,i\omega)\sim q^{2}$, we can obtain another
conclusion. We think that this case only takes place in a very low energy
range, although we cannot give an exact character energy. In this case,
the CS term will provide an energy gap $\Delta$ to the transverse parts of
the gauge field and CS gauge field, which will drastically influence the
physical property of the system. The effective action (28) can be written as
\begin{equation}\begin{array}{rl}
S_{eff.}=& \displaystyle{\sum_{q,\omega_{n}}(\chi_{F}q^{2}+\frac{\gamma_{1}
|\omega_{n}|}{q}+\Delta)(\delta_{ij}-\frac{q_{i}q_{j}}{q^{2}})}
A_{i}(q,i\omega_{n})A_{j}(-q,-i\omega_{n})\\
+& \displaystyle{\sum_{q,\omega_{n}}(\chi q^{2}+\frac{\gamma_{2}
|\omega_{n}|}{q}+\Delta)(\delta_{ij}-\frac{q_{i}q_{j}}{q^{2}})}
a_{i}(q,i\omega_{n})a_{j}(-q,-i\omega_{n})\\
-& \displaystyle{\frac{\alpha}{4\pi}\sum_{q,\omega_{n}}}\epsilon_{\mu\nu
\lambda}A_{\mu}(q,i\omega_{n})q_{\nu}A_{\lambda}(-q,-i\omega_{n})
\end{array}\end{equation}
where in the mean field theory (RPA) approximation$^{[23]}$, $\Delta\sim
\frac{\delta}{m}$, $\delta$ is the holon density (doping density). As
$\Delta\ll\chi q^{2}_{0}$, or $\chi_{F}q^{'2}_{0}$, $q_{0}=(\gamma_{2}
\omega_{0}/\chi)^{1/3}$, $q^{'}_{0}=(\gamma_{1}\omega_{0}/\chi_{F})^{1/3}$,
$\omega_{0}$ is some characteristic energy in which the overdamped mode
is dominant, we can omit the gap $\Delta$ although it is relevant as
taking the scaling transformations (29), and obtain the same conclusion as
that there existing longitudinal screening effect of the CS gauge field.
As the gap $\Delta$ is dominant, we must take following scaling
transformations
\begin{equation}\begin{array}{rl}
q& \rightarrow sq, \;\;\;\; \omega_{n}\rightarrow s\omega_{n}\\
a_{i}& \rightarrow s^{-3/2}a_{i}, \;\;\;\; A_{i}\rightarrow s^{-3/2}
A_{i}, \;\;\; s\rightarrow 0.
\end{array}\end{equation}
which keeps the quadratic terms but the $q^{2}$ terms in (33) invariance.
We see that the dynamic exponent $z=1$.
However, under these transformations, the CS and $q^{2}$ terms are
irrelevant, $\alpha(s)\rightarrow 0, \chi_{F}(s)(\chi(s))\rightarrow 0$,
as $s\rightarrow 0$. At this quantum critical point, the effective action
(33) can be written as
\begin{equation}\begin{array}{rl}
S^{c}_{eff.}=& \displaystyle{\sum_{q,\omega_{n}}(\frac{\gamma_{1}
|\omega_{n}|}{q}+\Delta)(\delta_{ij}-\frac{q_{i}q_{j}}{q^{2}})}
A_{i}(q,i\omega_{n})A_{j}(-q,-i\omega_{n})\\
+& \displaystyle{\sum_{q,\omega_{n}}(\frac{\gamma_{2}|\omega_{n}|}
{q}+\Delta)(\delta_{ij}-\frac{q_{i}q_{j}}{q^{2}})}
a_{i}(q,i\omega_{n})a_{j}(-q,-i\omega_{n})
\end{array}\end{equation}
By using these gauge propagators,
we can obtain the imaginary part of the fermion self-energy
\begin{equation}
\Sigma^{"}_{c}(k_{F}, \omega)\sim{\omega}^{2}
\end{equation}
We see that at this quantum critical point, the system has the Fermi
liquid behavior.

We now consider the transport properties of the system. We first assume that
there exists an energy scale $\omega_{c}$ which will be used to characterize
the nature of the hard-core bosons. If the energy $\omega<\omega_{c}$,
the nature of the hard-core bosons is important, i.e., the fermionic character
of the hard-core bosons is important. If the energy $\omega>\omega_{c}$,
the nature of the hard-core bosons can be neglected, i.e., the bosonic
character of the hard-core bosons is dominant.
In the case, we must use the action (27) to determine the temperature
dependence of the resistivity, it is well known$^{[12,13]}$
$\rho_{B}(T)\sim T$. In the case of
$\omega<\omega_{c}$,
if there exists the longitudinal screening effect of the CS gauge field,
the CS term has less influence on the transverse parts of the gauge field
and CS gauge field,
according to equation (32), the temperature dependence
of the resistivity is$^{[12]}$ $\rho_{B}(T)\sim T^{4/3}$, the system has a
non-Fermi
liquid behavior. If there is not the longitudinal screening effect of the
CS gauge field, the CS term has drastically influence on the transverse
parts of the gauge field and CS gauge field and provides an energy gap to
them, according to equation (36), the temperature dependence of the
resistivity is $\rho_{B}(T)\sim T^{2}$, the system has a Fermi liquid
behavior.
So we have the relation
\begin{equation}
\rho_{B}(T)\propto\left\{\begin{array}{ll}
T, & \mbox{$T>\omega_{c}$}\\
T^{\alpha^{'}}, & \mbox{$T<\omega_{c}$}
\end{array}\right.
\end{equation}
where $\alpha^{'}=4/3\;or\;2$. In fact, we have three characteristic
energies $T^{*}, \omega_{c} \; and\; \bar{T}^{*}$ to indicating the
behavior of the system, here $\bar{T}^{*}$ is the characteristic energy
indicating the crossover from $z=2$ to $z=1$ QC regimes for the spin
part. Only under the condition $T>max\{\omega_{c}, \bar{T}^{*}\}$, the
physical resistivity satisfies the relation $\rho(T)\propto T$. However,
in the little doping case, $\bar{T}^{*}\rightarrow\infty$, we can have
the relation $\rho(T)\propto T \;for\;T>max\{\omega_{c}, T^{*}\}$. From
the equation (23) we see that the characteristic energy $\bar{T}^{*}$
corresponds to the temperature at which the static spin susceptibility
tends to its maximal value as the system goes from the $z=1$ QC regime to
the $z=2$ QC regime, so $\bar{T}^{*}$ should be similar as the
characteristic temperature $T_{\chi max}$ defined in Ref.[31,32], at
which the spin susceptibility exhibits a broad peak. Generally, we have
the condition $\bar{T}^{*}<T_{\chi max}$, because there may exist a
transition region from $\chi(q=0)\propto\sqrt{T}$ to $\chi(q=0)\propto
const.$ in the $z=2$ QC regime which is dominant. The size of the
transition region depends upon the shape of the Fermi surface of the
doping holes. The experimental data in Ref.32 supports the above analysis
for the $La_{2-\delta}Sr_{\delta}CuO_{4}$ sample in the doping range
$0.09<\delta<0.16$, $\bar{T}^{*}\leq T_{\chi max}$, if in this doping
range the characteristic energy $\omega_{c}$ is less than $\bar{T}^{*}$.

We now consider the behavior of the temperature dependence of the Hall
coefficient.
We see that as there existing the longitudinal screening effect of the CS
gauge field, the parity-odd
gauge interaction has a little influence on the resistivity,
but it drastically influences the Hall coefficient. An external magnetic
field may have an activation effect on the nature of the hard-core bosons
and turns on the parity-odd gauge interaction $D_{jo}(q,\omega)=<a_{j}a_{0}>
=\sigma\varepsilon_{ij}q_{i}F(q^{2},\omega), F(q^{2},\omega)=<a_{i}a_{i}>
<a_{0}a_{0}>$. This parity-odd gauge interaction will exist in a wide
broad energy scale even to a higher energy in which the nature of the
hard-core boson is less important because of the statistical parameter
$\alpha$
being exactly marginal. If we take $F(q^{2},\omega)=\frac{1}{\epsilon_{0}}
\frac{1}{q^{2}}\delta_{\omega,0}$, using the method in Ref.33 to
calculate the Hall coefficient at lowest-order approximation, we get
\begin{equation}
R_{H}=R^{*}_{H}\frac{1}{1+\beta T}+R^{\infty}_{H}
\end{equation}
where $\beta=\frac{1}{8\pi^{2}\epsilon_{0}}(n_{F}-n_{F}^{2}),
n_{F}=\{exp(-\mu/k_{B}T)+1\}^{-1}$ is the Fermi distribution function,
the chemical potential $\mu$ is proportional to the doping density
$\delta$. At the low doping and high temperature, the coefficient
$\beta$ is nearly independent of temperature. $R^{*}_{H}$ is the Hall
coefficient for the system without parity-odd gauge interaction,
at low doping $R^{*}_{H}\propto\frac{1}{\delta}$, $R^{\infty}_{H}$
the Hall coefficient in the infinite temperature limit which can
be identified$^{[34]}$ $R^{\infty}_{H}=R_{0}(\frac{1}{4\delta}-
\frac{1}{1-\delta}+\frac{3}{4})$, $R_{0}$ is a constant.
We see that the anomalous temperature dependence of the Hall coefficient
derives from the parity-odd gauge interaction which is the respondence
of the nature of the hard-core bosons. The expression of the Hall
coefficient $R_{H}$ in (38) can better explain the current experimental
data$^{[35,36]}$.

In sumarry, we have used a simple physical picture that the doping holes
with single occupation constraint move in the antiferromagnetic
background of the copper spins to describe the normal state of the
cuprate superconducting materials by the traditional slave boson method.
We have classified the first Brillouin zone of the copper spins into as
two regions, one is the central region (near $q\sim 0$), which
controlls the behavior of the charge part of the doping holes and determines
the transport property of the system;  Another region is the corner
region, i.e., near the regions ${\bf
Q}=(\pm\frac{\pi}{a},\pm\frac{\pi}{a})$,  which controlls the
behavior of the copper and doping spins and determines the magnetic
property of the system. For the spin part, taking the long range
antiferromagnetic N\'{e}el order of the copper spins as a background and
using the renormalization method, we have given the spin susceptibilities
of the system in the $z=1$ and $z=2$ regimes, and used it to calculate
the NMR spin-lattice relaxation rate and spin echo rate. The results we
have obtained are in good agreement with the current experimental data.
For the charge part,
we have extensively studied a hard-core boson system
interacting via exchanging gauge bosons using the renormalization group
method. As considering the longitudinal screening of the CS gauge field,
the low energy and long wavelength behavior of the gauge
fields is shown to be in the Gaussian universality class with a dynamical
exponent z=3, and the parity-odd gauge interaction is exactly marginal.
The anomalous transport properties of cuprate materials is controlled
by the Hamiltonian at this quantum critical point $g(s)=g(0)$ and
$\alpha(s)=\alpha(0)$, the system has a non-Fermi liquid behavior. As
there is not the longitudinal screening effect of the CS gauge field, the
system has a Fermi liquid behavior.
We have showed that the physical resistivity is determined by both the
quasiparticle-spin-fluctuation and quasiparticle-gauge-fluctuation
scatterings, and the Hall coefficient is determined by the parity-odd
gauge interaction which derives from the nature of the hard-core bosons.

The author is very grateful to Prof. Z. B. Su and Prof. L. Yu for their
encouragement.

\newpage
{\bf References}
\bigskip
\begin{description}
\item [1]   J.G.Bednorz, and K.A.M\"{u}ller, Z. Phys. {\bf B}64, 189(1986).
\item [2]   P.W.Anderson, Science 235, 1196(1987).
\item [3]   F.C.Zhang, and T.M.Rice, Phys. Rev. {\bf B}37, 3759(1988).
\item [4]   V.J.Emery, Phys. Rev. Lett. {\bf 58}, 2794(1987);
P.B.Littlewood, C.M.Varma, and E.Abrahams, {\it ibid}, {\bf 60}, 379(1988).
\item [5]   C.M.Varma, P.B.Littlewood, S.Schmitt-Rink, E.Abrahams, and
A.Ruckenstein, Phys. Rev. Lett. {\bf 63}, 1996(1989).
\item [6]   A.Millis, H.Monien, and D.Pines, Phys. Rev. {\bf B}42,
167(1990); T.Moriga, Y.Takahashi, and K.Ueda, J. Phys. Soc. Jpn.
{\bf 59}, 2905(1990).
\item [7]   J.Zaanen, G.A.Sawatzky, and J.W.Allen, Phys. Rev. Lett. {\bf
55}, 418(1985).
\item [8]   S.Chakravarty, B.I.Halperin, and D.R.Nelson, Phys. Rev.
{\bf B}39, 2344(1989); S.Chakravarty, and R.Orbach, Phys. Rev. Lett. {\bf
64}, 224(1990).
\item [9]   S.Chakravarty, Proc. of 1989 Symposium on High-$T_{c}$
superconductivity, eds. K.S.Bedell, {\sl et. al.}, (Addison-Wesley, 1990)
and references therein.
\item [10]   S.Sachdev, and J.Ye, Phys. Rev. Lett. {\bf 69}, 2411(1992);
A.V.Chubukov, and S.Sachdev, Phys. Rev. Lett. {\bf 71}, 169(1993);
A.V.Chubukov, S.Sachdev, and J.Ye, Phys. Rev. {\bf B}49, 11919(1994).
\item [11]   L.Ioffe, and A.Larkin, Phys. Rev. {\bf B}39, 8988(1989);
P.A.Lee, Phys. Rev. Lett. {\bf 63}, 6801(1989).
\item [12]   N.Nagaosa, and P.A.Lee, Phys. Rev. Lett. {\bf 64}, 2450(1990);
Phys. Rev. {\bf B}43, 1223(1991); {\bf B}43, 1234(1991).
\item [13]   L.Ioffe, and P.W.Wiegmann, Phys. Rev. Lett. {\bf 65}, 653(1990);
L.B.Ioffe, and G.Kotliar, Phys. Rev. {\bf B}42, 10348(1990).
\item [14]  A.Sokol, and D.Pines, Phys. Rev. Lett. {\bf 71}, 2813(1993).
\item [15]  T.Imai, C.P.Slichter, A.P.Paulikas, and B.Veal,
Phys. Rev. {\bf B}47, 9158(1993).
\item [16]  T.Imai, C.P.Slichter, K.Yoshimura, and K.Kosuge,
Phys. Rev. Lett. {\bf 70}, 1002(1993).
\item [17]  T.Imai, C.P.Slichter, K.Yoshimura, M.Katoh, and K.Kosuge,
Phys. Rev. Lett. {\bf 71}, 1254(1993).
\item [18]  M.Takigawa, Phys. Rev. {\bf B}49, 4158(1994)
\item [19]  J.Rossat-Mignod, {\it et al.}, Physica (Amsterdam) 185-189C,
86(1991); P.M.Gehring, {\it et al.}, Phys. Rev. {\bf B}44, 2811(1991).
\item [20]  Y.L.Liu, J. Phys.: Condens. Matter, {\bf 7}, 3749(1995);
preprint.
\item [21]  Y.L.Liu, and Z.B.Su, Phys. Lett. {\bf A}200, 393(1995).
\item [22]  E.Fradkin, Phys. Rev. Lett. {\bf 63}, 322(1989); A.A.Ovchinnikov,
and An.A.Ovchinnikov, Mod. Phys. Lett. {\bf B}6, 1951(1992).
\item [23]  Y.L.Liu, and Z.B.Su, Phys. Rev. {\bf B}48, 13018(1993).
\item [24]  J.A.Hertz, Phys. Rev. {\bf B}14, 1165(1976);
D.R.Nelson, and R.A.Pelcovits, {\it ibid}, {\bf B}16, 2191(1977);
A.J.Millis, {\it ibid}, {\bf B}48, 7183(1993).
\item [25]  P.Monthoux, and D.Pines, Phys. Rev. {\bf B}50, 16015(1994);
V.Barzykin, and D.Pines, preprint.
\item [26]  J.Polchinski, in {\it Recent Direction in Particle Theory},
Proc. 1992 TASI, eds. J.Harvey and J.Polchinski; Nucl. Phys. {\bf B}422,
617(1994).
\item [27]  L.B.Ioffe, and A.J.Millis, Phys. Rev. {\bf B}51, 16151(1995);
S.Sachdev, A.V.Chubukov, and A.Sokol, {\it ibid}, {\bf B}51, 14874(1995).
\item [28]  J.Gan, and E.Wong, Phys. Rev. Lett. {\bf 71}, 4226(1993).
\item [29]  B.L.Altshuler, L.B.Ioffe, and A.J.Millis, Phys. Rev. {\bf
B}50, 14048(1994).
\item [30]  S.Chakravarty, R.E.Norton, and O.F.Syljuasen, Phys. Rev. Lett.
{\bf 74}, 1423(1995).
\item [31]  D.C.Johnston, Phys. Rev. Lett. {\bf 62}, 957(1989).
\item [32]  T.Nakano, {\it et al.}, Phys. Rev. {\bf B}49, 16000(1994).
\item [33]  L.B.Ioffe, V.Kalmeyer, and P.B.Wiegmann, Phys. Rev. {\bf B}43,
1219(1991).
\item [34]  B.Shastry, B.I.Shraiman, and R.R.P.Singh, Phys. Rev. Lett.
{\bf 70}, 2004(1993).
\item [35]  For a review see N.P.Ong, Physical Properties of High
Temperature Superconductors, edited by D.M.Ginsberg ( World Scientific,
Singapore, 1990), vol.2, p459.
\item [36]  H.Y.Huang, {\it et al.}, Phys. Rev. Lett. {\bf 72}, 2636(1994).

\end{description}
\end{document}